\documentclass[11pt]{article}
\usepackage{graphicx}
\usepackage{amsmath}
\usepackage{amssymb}
\usepackage{amsxtra}

\newcommand{\gsim}{\mathrel{\mathop{\kern 0pt \rlap
  {\raise.2ex\hbox{$>$}}}
  \lower.9ex\hbox{\kern-.190em $\sim$}}}

\title{DARK MATTER PARTICLES IN THE GALACTIC HALO}
\author{R. Bernabei\\Dip. di Fisica, Universit\`a di Roma ``Tor Vergata'' \\and INFN-Tor Vergata, I-00133 Rome, Italy}

\begin{document}
\maketitle

\begin{abstract}

The DAMA/LIBRA--phase1 and the former DAMA/NaI data
(cumulative exposure $1.33$ ton $\times$ yr, corresponding to 14 annual cycles)
give evidence at 9.3 $\sigma$ C.L.
for the presence of Dark Matter (DM)
particles in the galactic halo, on the basis of the exploited model independent
DM annual modulation signature by using highly radio-pure NaI(Tl) target.
Results and comparisons will be shortly addressed as well as perspectives of the presently running DAMA/LIBRA-phase2.
Finally, some arguments arisen in the discussion section
of this workshop are mentioned in the Appendix. 
\end{abstract}
\baselineskip=14pt

\section{Introduction}\label{s:intro}

About 80 years of experimental observations and theoretical arguments have pointed out 
that a large fraction of the Universe is composed by Dark Matter particles
\footnote{For completeness, it is worth recalling that some efforts to find alternative explanations to Dark Matter have been proposed
such as $MOdified$ $Gravity$ $Theory$ (MOG) and $MOdified$ $Newtonian$ 
$Dynamics$ (MOND); they hypothesize that the theory of gravity is incomplete and that a new gravitational theory might explain
the experimental observations.
MOND modifies the law of motion for very small accelerations, while MOG modifies the Einstein's theory of
gravitation to account for an hypothetical fifth fundamental force in addition to the gravitational, electromagnetic,
strong and weak ones. However, e.g.: i) there is no general underlying principle; ii)
they are generally unable to account for all small and large scale 
observations; iii) they fail to reproduce accurately the Bullet Cluster; iv) 
generally they require some amount of DM particles as seeds for the structure formation.}.

The presently running DAMA/LIBRA ($\simeq$ 250 kg of full sensitive target-mass)
\cite{perflibra,modlibra,bot11,pmts,mu,review,papep,diu2014,norole} experiment,
as well as the former DAMA\-/NaI ($\simeq$ 100 kg of full sensitive target-mass) \cite{allDM1,allDM,Nim98,RNC,ijma,chan,allRare},
has the main aim to investigate the presence of DM particles in the galactic halo by exploiting
the model independent DM annual modulation signature (originally suggested in Ref.~\cite{Freese}).

As a consequence of the Earth's revolution around the Sun,
which is moving in the Galaxy with respect to the Local Standard of
Rest towards the star Vega near
the constellation of Hercules, the Earth should be crossed
by a larger flux of DM particles around $\simeq$ 2 June
and by a smaller one around $\simeq$ 2 December.
In the former case the Earth orbital velocity is summed to the one of the
solar system with respect to the Galaxy, while in the latter
the two velocities are subtracted\footnote{Thus,
the DM annual modulation signature has a different origin and peculiarities
than the seasons on the Earth and than effects
correlated with seasons (consider the expected value of the
phase as well as other requirements listed below).}.
This DM annual modulation signature is very distinctive since the effect
induced by DM particles must simultaneously satisfy
all the following requirements: the rate must contain a component
modulated according to a cosine function (1) with
one year period (2) and a phase that peaks roughly
$\simeq$ 2 June (3); this modulation must only be found in a
well-defined low energy 
range, where DM particle induced
events can be present (4); it must apply only to those events
in which just one detector of many (9 in DAMA/NaI and 25 in DAMA/LIBRA) actually ``fires'' ({\it single-hit}
 events), since the DM particle multi-interaction probability
is negligible (5);
the modulation amplitude in the region
of maximal sensitivity must be $\simeq$  7\% for usually adopted
halo distributions (6), but it can be larger (even up to $\simeq$ 30\%)
in case of some
possible scenarios such as e.g. those in Ref.~\cite{Wei01,Fre04}.
Thus, this signature is model independent and very effective; moreover,
the developed highly radio-pure NaI(Tl) target-detectors \cite{perflibra} and the adopted procedures
assure sensitivity to a wide range
of DM candidates (both inducing nuclear recoils and/or electromagnetic 
radiation), interaction types and astrophysical scenarios.

In particular, the experimental observable in DAMA experiments is the modulated component of the signal in 
NaI(Tl) target and not the constant part 
of it as in other approaches as those by CDMS, Xenon, etc., 
where in addition e.g.: i) different target materials are used; ii) the sensitivity is mainly restricted to candidates inducing just nuclear recoils;
iii)  many (by the fact largely uncertain)
selections/subtractions of detectors and of data and (highly uncertain) extrapolations of detectors' features are applied. 

The DM annual modulation signature might be mimicked only by systematic effects or side reactions
able to account for the whole observed modulation amplitude and
to simultaneously satisfy all the requirements given above.
No one is available or suggested by anyone over more than a decade 
\cite{perflibra,modlibra,mu,review,diu2014,RNC,norole}. 

It is also worth noting that the DM annual modulation signature acts itself as a strong background reduction as pointed out since the early paper by Ref. \cite{Freese},
and especially when all the above peculiarities can be 
experimentally verified in suitable dedicated set-ups as it is the case of the DAMA experiments.

\section{The DAMA results}\label{s:res}

The total exposure of DAMA/LIBRA--phase1 is:
1.04 ton $\times$ yr in seven annual cycles;
when including also that of the first generation DAMA/NaI experiment it is
$1.33$ ton $\times$ yr, corresponding to 14 annual cycles.
The variance of the cosine during the DAMA/LIBRA--phase1 data taking is 0.518,
showing that the set-up has been operational evenly throughout the years
\cite{modlibra,review}.

\begin{figure*}[!t]
\begin{center}
\includegraphics[width=0.95\textwidth] {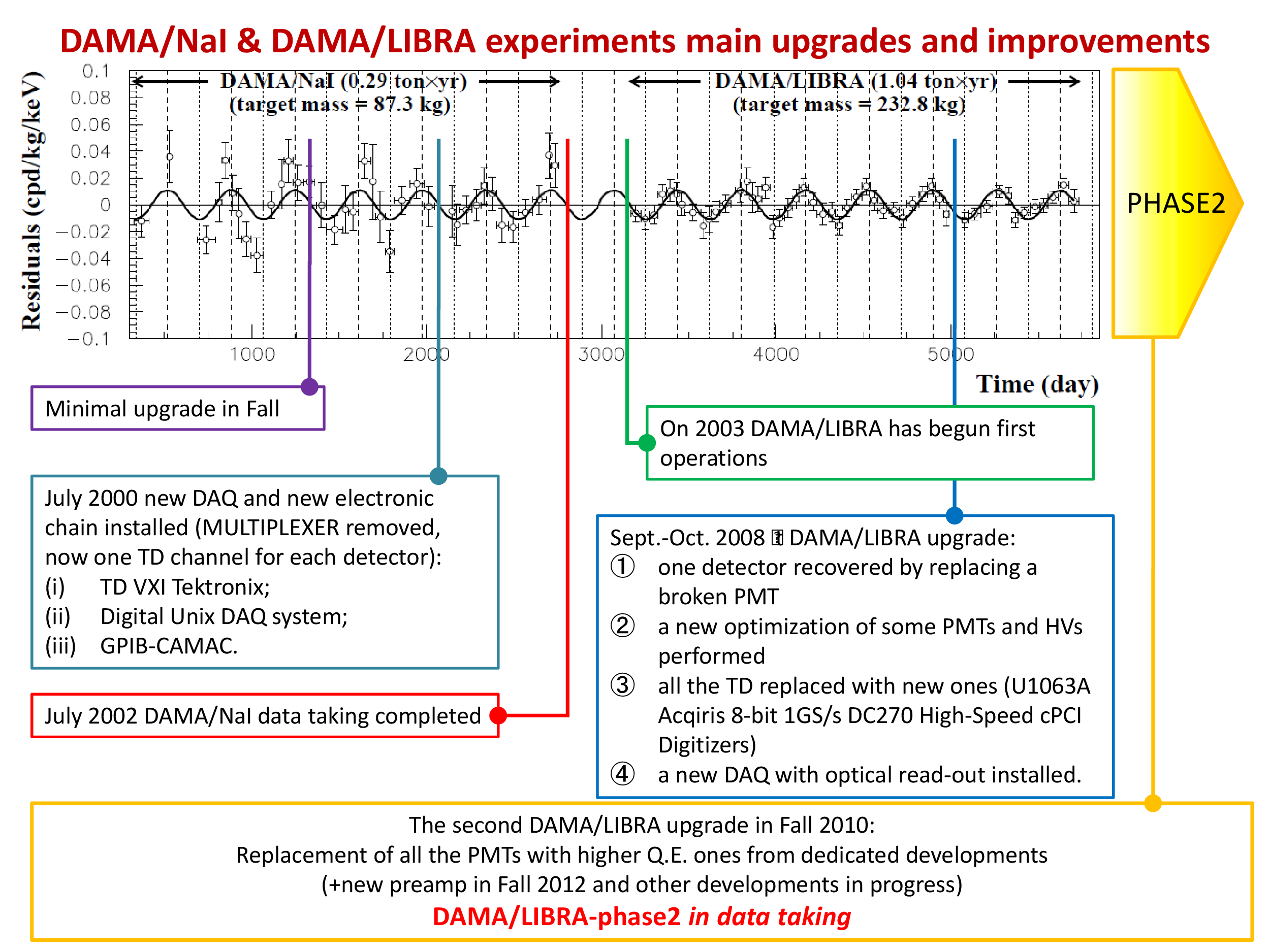}
\end{center}
\vspace{-.4cm}
\caption{Experimental residual rate of the {\it single-hit} scintillation events
measured by DAMA/NaI and DAMA/LIBRA--phase1 in the (2--6) keV energy interval
as a function of the time.
The data points present the experimental errors as vertical bars and the associated
time bin width as horizontal bars; see text. As always in DAMA results, the given rate is already corrected for the overall efficiency.
The major upgrades of the experiment are also pointed out.}
\label{fg:res}
\end{figure*}

Many independent data analyses have been carried out \cite{modlibra,review}
and all of them confirm the presence of a peculiar annual modulation in the {\it single-hit} scintillation events in the 
2-6 keV energy interval,
which -- in agreement with the requirements of the DM signature -- is absent in other parts of the energy spectrum and 
in the {\it multiple-hit} scintillation events in the same 2-6 keV energy interval (this latter condition 
correspond to have ``switched off the beam" of DM particles).  All  the analyses and details can be found in the literature given 
above. In particular, Fig.~\ref{fg:res} shows the time behaviour of the experimental
residual rates of the {\it single-hit} scintillation
events for DAMA/NaI \cite{RNC} and DAMA/LIBRA--phase1 \cite{modlibra,review} cumulatively in the (2--6) keV energy interval.
The data points present the experimental errors as vertical bars and the associated  
time bin width as horizontal bars.
The superimposed curve is the cosinusoidal function $A \cos \omega(t-t_0)$
with a period $T = \frac{2\pi}{\omega} =  1$ yr, a phase $t_0 = 152.5$ day (June 2$^{nd}$) and
modulation amplitude, $A$, equal to the central value obtained by best fit on the data points.
The dashed vertical lines
correspond to the maximum expected for the DM signal, while
the dotted vertical lines correspond to the expected minimum. The major upgrades are also pointed out.

In order to continuously monitor the running conditions, several pieces of information are acquired with the production data
and quantitatively analysed. In particular, all the time behaviours
of the running parameters, acquired with the production data,
have been investigated: the modulation amplitudes obtained for each
annual cycle when fitting the time behaviours of the parameters including a cosine
modulation with the same phase and period as for DM particles are well compatible with zero.
In particular, 
no modulation has been found in any
possible source of systematics or side reactions; thus, cautious upper limits (90\% C.L.)
on possible contributions to the DAMA/LIBRA measured modulation amplitude
have been derived (see e.g. \cite{modlibra}).
It is worth noting that they do not quantitatively account for the
measured modulation amplitudes, and are not able to simultaneously satisfy all the many requirements of the signature.
Similar analyses have also been carried out for
the DAMA/NaI data\cite{RNC}.

\begin{table*}[!ht]
\caption{Summary of the contributions to the total neutron flux at LNGS; the value, the relative modulation amplitude,
and the phase of each component is reported.
It is also reported the counting rate in DAMA/LIBRA for {\it single-hit}
events, in the ($2-6$) keV energy region induced by neutrons, muons and solar neutrinos, detailed for each component.
The modulation amplitudes, $A_k$, are reported as well, while the last column shows the relative contribution
to the annual modulation amplitude observed by DAMA, $S_m^{exp} \simeq 0.0112$ cpd/kg/keV \cite{modlibra}.
As can be seen, they are all negligible and they cannot give any significant contribution to
the observed modulation amplitude. In addition, neutrons, muons and solar neutrinos are not a competing background
when the DM annual modulation signature is investigated since in no case they can mimic this signature. For details
see Ref. \cite{norole} and references therein.}
\label{table:tab12}
\vspace{0.2cm}
\resizebox{0.98\textwidth}{!}{
\begin{tabular}{|ll|ccc|cc|c|}
\hline
\multicolumn{2}{|c|} {Source} & $\Phi^{(n)}_{0,k}$                & $\eta_k$                                    & $t_k$ & $ R_{0,k}$   & $A_k = R_{0,k} \eta_k $        & $A_k/S_m^{exp}$ \\
 &                     & (neutrons cm$^{-2}$ s$^{-1}$)            &                                             &       & (cpd/kg/keV) & (cpd/kg/keV) & \\
\hline
 & thermal n & $1.08 \times 10^{-6}$     & $\simeq 0$                                  & --    & $<8 \times 10^{-6}$ & $\ll 8 \times 10^{-7}$ & $\ll 7 \times 10^{-5}$ \\
 & ($10^{-2}-10^{-1}$ eV) &                          & however $\ll 0.1$  &       &   & & \\
 SLOW & & & & & & & \\
 neutrons & epithermal n & $2 \times 10^{-6}$           & $\simeq 0$                                  & --    &   $<3 \times 10^{-3}$ & $\ll 3 \times 10^{-4}$ & $\ll 0.03$ \\
 & (eV-keV)     &                                          & however $\ll 0.1$  &       &   &   & \\
\hline
 & fission, $(\alpha,n) \rightarrow$ n  & $\simeq 0.9 \times 10^{-7}$  & $\simeq 0$                                  & --  & $< 6 \times 10^{-4}$ & $\ll 6 \times 10^{-5}$ & $\ll 5 \times 
10^{-3}$ \\
 & (1-10 MeV)               &                                          & however $\ll 0.1$  &     &   &   & \\
 & & & & & & & \\
     & $\mu \rightarrow $ n from rock    & $\simeq 3 \times 10^{-9}$                & 0.0129                      & end of  & $\ll 7 \times 10^{-4}$ & $\ll 9 \times 10^{-6}$ & $\ll 8 \times 
10^{-4}$ \\
 FAST  & ($> 10$ MeV)             &          &                                             &                        June    &                        &  & \\
 neutrons & & & & & & & \\
     & $\mu \rightarrow $ n from Pb shield & $\simeq 6 \times 10^{-9}$                & 0.0129                      & end of  & $\ll 1.4 \times 10^{-3}$ & $\ll 2 \times 10^{-5}$ & $\ll 1.6 
\times 10^{-3}$ \\
     & ($> 10$ MeV)             &         &                                             &                             June    &                        &   & \\
 & & & & & & & \\
 & $\nu \rightarrow $ n     & $\simeq 3 \times 10^{-10}$   & 0.03342$^*$                               & Jan. 4th$^*$ & $\ll 7 \times 10^{-5}$ & $\ll 2 \times 10^{-6}$ & $\ll 2 \times 10^{-4}$ 
\\
 & (few MeV)                & & &   &   &   & \\
\hline
\multicolumn{2}{|c|} {direct $\mu$}       & $\Phi^{(\mu)}_{0} \simeq 20$ $\mu$ m$^{-2}$d$^{-1}$   & 0.0129  & end of  & $\simeq 10^{-7}$ & $\simeq 10^{-9}$ & $\simeq 10^{-7}$ \\
 & & & & June & & & \\
\multicolumn{2}{|c|} {direct $\nu$}       & $\Phi^{(\nu)}_{0} \simeq 6 \times 10^{10}$ $\nu$ cm$^{-2}$s$^{-1}$  & 0.03342$^*$          & Jan. 4th$^*$                         & $\simeq 10^{-5}$ 
& $3 \times 10^{-7}$ & $3 \times 10^{-5}$ \\
\hline
\end{tabular}}
\vspace{0.3cm}
\footnotesize{\\$^*$ The annual modulation of solar neutrino is due to the different Sun-Earth distance along the year; so the relative modulation amplitude is twice the
eccentricity of the Earth orbit and the phase is given by the perihelion.}
\end{table*}
%

No other experimental result has been verified over so long time
so accurately and with various significant upgrades of the set-ups.

For completeness I mention that sometimes naive statements
were put forwards as the fact that
in nature several phenomena may show some kind of periodicity.
The point is whether they could
mimic the annual modulation signature in DAMA/LIBRA (and former DAMA/NaI), i.e.~whether they
could quantitatively account for the observed
modulation amplitude and also simultaneously
satisfy all the requirements of the DM annual modulation signature. The same is also for side reactions.
This has already been deeply investigated in Ref.~\cite{perflibra,modlibra} and references
therein;
the arguments and the quantitative conclusions, presented there, also
apply to the entire DAMA/LIBRA--phase1 data. Additional arguments can be found
in Ref.~\cite{mu,review,diu2014,norole}.
In particular, Ref. \cite{norole} further outlines in a simple and intuitive way why neutrons (of whatever origin), muons and 
solar neutrinos cannot give any significant contribution to the DAMA annual modulation results and -- in addition -- can never mimic the DM annual modulation signature since some of its specific requirements fail. 
Table \ref{table:tab12} summarizes the safety upper limits on the contributions (if any) to the observed modulation amplitude
due to the total neutron flux at LNGS,
either from $(\alpha$,n) reactions, from fissions and from muons' and solar-neutrinos' interactions in the rocks and in the lead
around the experimental set-up; the direct contributions of muons and solar neutrinos are also reported there.
As seen in Table \ref{table:tab12}, they are all negligible and they cannot give any significant contribution to
the observed modulation amplitude; in addition, neutrons, muons and solar neutrinos are not a competing background
when the DM annual modulation signature is investigated since they cannot mimic this signature. For details
see Ref. \cite{norole} and references therein.

In conclusion, DAMA gives a model-independent evidence -- at 9.3$\sigma$ C.L. over 14 independent annual cycles --
for the presence of DM particles in the galactic halo.
 
\subsection{On comparisons}\label{s:sub1}

No direct model independent comparison is possible in the field when different target materials 
and/or approaches are used; the same is for the strongly model dependent indirect searches\footnote{It should be noted that 
the rising behaviour of the positron flux reported in Ref. \cite{pamela,ams2} does not give any intrinsic evidence for production 
due to DM annihilation; 
this may arise only when a particular model of the competing background is assumed as e.g. the 
GALPROP code. But other more complete models 
exist which do not support any significant excess evidence. Moreover, an interpretation in terms of DM particle annihilation would require the assumption of: 
i) a very large boost factor ($\sim$ 400) of the density; ii) to boost the annihilation cross section through an assumed new interaction type; iii) to adjust the propagation
parameters; iv) to consider extra-source (subhalos, IMBHs); v) to consider only a leptophilic candidate to justify the absence of any excess in the 
antiproton spectrum. 
Finally, other well known sources can account for a similar
positron fraction: pulsars, supernova
explosions near the Earth, SNR.}.

In order to perform corollary investigations on the nature of the DM particles, model-dependent
analyses are necessary\footnote{For completeness, it is worth recalling that it does not exist any approach to investigate the nature 
of the candidate in the direct and indirect DM searches, which can offer this information independently on 
assumed astrophysical, nuclear and particle Physics scenarios. On the other hand, searches for new particles beyond the
Standard Model of particle Physics at accelerators cannot
credit by themselves that a certain particle is in the halo as a solution or the only solution for DM particles, and -- in addition --
DM candidates and scenarios (even for the neutralino) exist which cannot be investigated there.}. 
Thus, many theoretical and experimental parameters and 
models are possible (see e.g. in \cite{modlibra,review,norm,Wall14}) and many hypotheses must also be exploited, while specific experimental and theoretical assumptions are generally 
adopted in the field assuming a single arbitrary scenario without accounting neither for existing uncertainties nor for 
alternative possible scenarios, interaction types, etc.

The obtained DAMA 9.3 $\sigma$ C.L. model independent evidence
is compatible with a wide set of scenarios regarding the nature of the DM candidate
and related astrophysical, nuclear and particle Physics. For examples
some scenarios and parameters are discussed e.g. in
Ref.~\cite{allDM1,allDM,RNC,modlibra,review,norm,Wall14}.
Further large literature is available on the topics (see for example in the bibliography of Ref. \cite{review}).
By the fact, both the negative results and all the possible positive hints
are largely compatible with the DAMA model-independent DM annual
modulation results in various scenarios considering also the existing experimental and
theoretical uncertainties; the same holds for the strongly model dependent indirect approaches.

It is also worthwhile to further recall that these DAMA experiments are not only sensitive to DM particles with spin-independent coupling 
inducing just nuclear recoils, but also to other couplings and to other DM candidates 
as those giving rise to part or all the signal in electromagnetic form. Finally, scenarios exist in which 
other kind of targets/approaches are disfavoured or even blind.

\section{DAMA/LIBRA--phase2 and perspectives}\label{s:ph2}

An important upgrade has started at end of 2010 replacing all the PMTs with new ones having higher Quantum Efficiency;
details on the developments and on the reached performances in the operative conditions
are reported in Ref. \cite{pmts}. They have allowed us to lower the software energy threshold of the experiment 
to 1 keV and to improve also other features as e.g. the energy resolution \cite{pmts}.

Since the fulfillment of this upgrade and after some optimization periods, DAMA/LIBRA--phase2
is continuously running in order e.g.:
(1) to increase the experimental sensitivity thanks to the lower software energy threshold; 
(2) to improve the corollary investigation on the nature of the
DM particle and related astrophysical, nuclear and particle physics arguments;
(3) to investigate other signal features and second order effects. This requires long and
 dedicated work for reliable collection and analysis of very large
exposures.

In the future DAMA/LIBRA will also continue its study on several other rare
processes as also the former DAMA/NaI apparatus did.

Finally, further future improvements of the DAMA/LIBRA set-up to increase the sensitivity (possible DAMA/LIBRA-phase3) and the 
developments towards the possible DAMA/1ton (1 ton full sensitive mass on the contrary of other kind of detectors), we proposed in 1996, 
are considered at some extent. For the first case developments of new further radiopurer PMTs with high quantum efficiency are starting,
while in the second case it would be necessary to overcome the present problems regarding: i) the supplying, 
selection and purifications of a large number of high quality NaI and, mainly, TlI powders; ii) the 
availability of equipments and competence for reliable measurements of small trace contaminants in ppt or lower region; 
iii) the creation 
of updated protocols for growing, handling 
and maintaining the crystals; iv) the availability of large Kyropoulos equipments 
with suitable platinum crucibles; v) etc.. At present, due to the change of rules for provisions of strategical materials, 
the large costs and the lost of some equipments and competence also at industry level, a 
satisfactory development appears quite difficult.

\section{Conclusions}\label{s:concl}

The data of DAMA/LIBRA--phase1 have further confirmed the presence of a
peculiar annual modulation of the {\it single-hit} events in the (2--6) keV energy region
satisfying all the many requirements of the DM annual modulation signature;
the cumulative
exposure by the former DAMA/NaI and DAMA/LIBRA--phase1 is
1.33 ton $\times$ yr (orders of magnitude larger than those typically released in the field).

As required by the DM annual modulation signature:
1) the {\it single-hit} events show a clear cosine-like modulation as expected for the DM signal;
2) the measured period is equal to $(0.998\pm 0.002)$ yr well compatible with the 1 yr period as expected for the DM signal;
3) the measured phase $(144\pm 7)$ days is compatible with $\simeq$ 152.5 days as expected for the DM signal;
4) the modulation is present only in the low energy (2--6) keV interval and not in other higher energy regions, consistently with expectation for the DM signal;
5) the modulation is present only in the {\it single-hit} events, while it is absent in the {\it multiple-hit} ones as expected for the DM signal;
6) the measured modulation amplitude in NaI(Tl) of the {\it single-hit} events in the (2--6) keV energy
   interval is: $(0.0112 \pm 0.0012)$ cpd/kg/keV (9.3 $\sigma$ C.L.).
No systematic or side processes able to simultaneously satisfy all the many
peculiarities of the signature and to account for the whole measured modulation  
amplitude is available.

DAMA/LIBRA--phase2 is continuously running in its new configuration with a lower software energy threshold aiming to improve
the knowledge on corollary aspects regarding the signal and on second order effects as discussed e.g. in Ref.~\cite{review,diu2014}.

Few comments on model--dependent comparisons have also been addressed here.

\section*{Acknowledgments}

It is a pleasure to thank all my DAMA collaborators who effectively dedicated their efforts to this experimental activity
and the colleagues in this Workshop for the interesting topics we have discussed, for the question 
section, and for the pleasant scientific environment.

\section*{Appendix: Questions \& Answers}

This section shortly summarizes some of the topics extensively discussed at the Workshop, where the time dedicated to discussions and the interest in deeply
understanding the topics were rather large.

\vspace{0.4cm}

 {\it Question 1: may you comment about the ratio of the measured dark matter particles 
    modulation amplitude to the total signal:  the $S_m/S_0$ ratio?}

\vspace{0.3cm}

Answer 1: the measured counting rate in the cumulative energy spectrum is about 1 cpd/kg/keV in the lowest energy bins; this  
is the sum of the background contribution and of the constant part of the signal S$_0$. As discussed e.g. in 
TAUP2011 \cite{taupnoz}, the background in the 2-4 keV energy region is estimated to be 
not lower than about 0.75 cpd/kg/keV; this gives an upper limit on S$_0$ of about 0.25 cpd/kg/keV. Thus, the S$_m$/S$_0$ ratio is equal or larger than 
about 0.01/0.25 $\simeq$ 4 \%.

\vspace{0.4cm}

 {\it Question 2: may you comment on the quenching factors, on their dependence on the type of the particles, and on 
       some typical examples of extreme properties?}

\vspace{0.3cm}

 Answer 2:  The quenching factor values play a role only when corollary model-dependent analyses for DM candidates inducing just nuclear recoils are carried out,
in order to derive the energy scale in terms of nuclear recoil energy.

As is widely known, the quenching factor is a specific property of 
the employed detector and not a general quantity universal 
for a given material. For example, in liquid noble-gas detectors, it depends -- among others -- on the 
level of trace contaminants which can vary in time and from one liquefaction process to another, on the cryogenic microscopic conditions, etc..
In bolometers it depends for instance on specific properties, trace contaminants, cryogenic conditions, 
etc. of each specific detector, while generally it is assumed exactly equal to unity (the maximum possible value). 
The quenching factors in scintillators depend, for example, on the dopant concentration, on the growing method/procedures, on residual trace contaminants, 
etc., 
and are expected to be energy dependent. Thus, all these aspects are already by themselves relevant sources of uncertainties 
when interpreting whatever result in terms of DM candidates inducing just nuclear recoils. 
Similar arguments have been addressed e.g. in Ref. \cite{modlibra,bot11,RNC,chan,tretyak}. 

\vspace{0.4cm}

 {\it Question 3: May you comment under which extreme conditions your experiment is successful and
    comment what can at most the experiment which does not fulfil one of the
    conditions or more than one of them at most can ``see"?}

\vspace{0.3cm}

 Answer 3: The full description and potentiality of the DAMA/LIBRA set-up have been discussed in details in Refs. \cite{perflibra,modlibra,pmts}
and references therein. Obviously all the set-up specific features and adopted procedures contribute to the possibility to point out the signal
through the model independent DM annual modulation signature. 
The absence/difference of one of them would limit whatever else result. 

\vspace{0.4cm}                                                                   

 {\it Question 4: May you comment about muons?}

 \vspace{0.3cm}                                                                   

Answer 4: An extensive discussion on this topics can be found in the dedicated Ref. \cite{mu,norole}, where its has been quantitatively demonstrated 
(see also Table \ref{table:tab12} in this paper) that -- for many reasons (and just one would suffice) - muons cannot play (directly or indirectly) 
any role in the DAMA annual modulation effect.

\vspace{0.4cm}
 {\it Question 5: May you comment about neutrinos?}
\vspace{0.3cm}

Answer 5: The contribution from solar, atmospheric, .. neutrinos is obviously negligible; a quantitative discussion can be found in Ref. \cite{norole}
(see also Table \ref{table:tab12} in this paper).  

\vspace{0.4cm}

 {\it Question 6:  May you comment about the operating temperature of your measuring apparatus?}

 \vspace{0.3cm}                                                                   

Answer 6: The DAMA set-ups operate at environmental temperature maintained stable by suitable and redundant air-conditioning system 
(2 independent devices for redundancy); moreover, the Cu housings of the detectors are in direct contact with the multi-ton metallic shield, thus 
there is a huge heat capacity ($\sim$ 10$^6$ cal/$^0$C). In addition, the operating temperature of the detectors is continuously 
monitored and analysed as the production data. A discussion on temperature in 
operating condition can be found e.g. in Ref. \cite{modlibra,review}.

\vspace{0.4cm} 
 
 {\it Question 7:  May you comment about the Snowmass plots and its  meaning?}
 \vspace{0.3cm}

Answer 7:
The recent plot from Snowmass and that in Ref. \cite{RPP}
about the ``status of the Dark Matter search"
do not point out at all the real status of Dark Matter searches since e.g.: i) Dark Matter has 
wider possibilities than WIMPs inducing just nuclear recoil with spin-independent interaction under single (largely 
arbitrary) set of assumptions; 
ii) neither the uncertainties for existing experimental and theoretical aspects nor alternative possible assumptions 
are accounted for; iii) they do not include possible systematic errors affecting the data 
(such as e.g. ``extrapolations" of energy threshold, of energy resolution and of
efficiencies, quenching factors values, 
convolution with poor energy resolution, correction for non-uniformity of the detector, multiple subtractions/selection of 
detectors and/or data, 
assumptions on quantities related to halo model, form factors, scaling laws, etc.); 
iv) the DAMA implications -- even adopting the many arbitrary assumptions considered there -- appear incorrect, for example the 
S$_0$ prior is not accounted for, etc., etc.. The perspectives as well appear incorrect/too optimistic.

\end{document}